\documentclass{article}
\pdfoutput=1
\usepackage[utf8]{inputenc}
\usepackage{authblk}
\usepackage{setspace}
\usepackage[margin=1.25in]{geometry}
\usepackage{graphicx}
\graphicspath{ {./figures/} }
\usepackage{subcaption}
\usepackage{amsmath}
\usepackage{lineno}
\usepackage{lmodern}
\usepackage{autobreak}%公式自动换行
\usepackage{booktabs} % 为tabular提供更好看的hline命令
\usepackage{array}% tabular using m{2cm}, m , vertical alignment
\usepackage{mathrsfs}%使用花体字 命令\mathscr
\usepackage{physics}
\usepackage{xparse}

\usepackage{geometry}%页边距
\usepackage[bookmarks=true,colorlinks=true,citecolor=blue]{hyperref} %超链接
\hypersetup{linkcolor=[rgb]{0,0,0}, bookmarksopen = true}% 更多设置请查阅: texdoc hyperref
\usepackage{float}
\usepackage[]{tikz}
\usepackage[]{pgfplots}
\pgfplotsset{compat=newest}
\usetikzlibrary{positioning}   % tikz绘图相对位置
\tikzset{elegant/.style={smooth,thick,samples=50,cyan}}
\numberwithin{equation}{section}
\usepackage[inline]{enumitem} % 编号
\allowdisplaybreaks
\usepackage{titlesec}
\titleformat{\section}{\fontsize{16}{19.2}\bfseries}{\thesection}{1em}{}
\titleformat{\subsection}{\fontsize{14}{16.8}\bfseries}{\thesubsection}{1em}{}

\usepackage{multirow}

\title{\Large{\textbf{Islands on codim-2 branes in Gauss-Bonnet Gravity}}}

\author{\small Zhengjiang Li\footnote{lizhj66@mail2.sysu.edu.cn} 
 }
\author{\small Zekai Hong\footnote{hongzk@mail2.sysu.edu.cn}}

\affil{\small{\textit{School of Physics and Astronomy, Sun Yat-Sen University},\\\textit{2 Daxue Road, Zhuhai 519082, China}}}
\date{}

\begin{document}

\maketitle

%%%%%% Abstract %%%%%%
\begin{abstract}
We study the black hole information problem on codim-2 branes in Gauss-Bonnet gravity. Thanks to the island surface ending on the brane, the Page curve of eternal black holes can be recovered for all of the GB couplings within the causal constraints. Our results strongly support the universality of the island mechanism.  Similar to Einstein's gravity, the HM surface can exist only in a finite time in GB gravity. Remarkably, for various parameters, the maximum times of HM surface are always larger than the Page times. As a result, the strange behavior of HM surfaces does not affect the Page curves for general GB gravity. 
%Finally, we find that the Page time increases with the GB couplings and the brane tension. 
Finally, we establish the correlation between the Page time, GB couplings, and brane tension, revealing that the Page time increases with these factors.
\end{abstract}

\tableofcontents
%%%%%% Main Text %%%%%%

\section{Introduction}

Double holography plays a vital role in addressing the black hole information paradox \cite{Hawking:1976ra, Penington:2019npb, Almheiri:2019psf, Almheiri:2019hni, Almheiri:2020cfm}. The most studied doubly holographic modes are 
Karch-Randall (KR) braneworld \cite{Karch:2000ct}, AdS/BCFT correspondence \cite{Takayanagi:2011zk, Miao:2017gyt, Chu:2017aab, Miao:2018qkc, Chu:2021mvq} and wedge holography \cite{Akal:2020wfl, Miao:2020oey,Hu:2022lxl}, where the typical brane is codim-1. See also \cite{Almheiri:2019yqk,Almheiri:2019psy,Geng:2020qvw,Chen:2020uac,Chen:2020hmv,Ling:2020laa,Krishnan:2020fer,Miao:2022mdx,Miao:2023unv,Li:2023fly,Chou:2021boq,Ahn:2021chg,Alishahiha:2020qza,Hu:2022ymx,Chu:2022ieq,Yu:2022xlh,Geng:2020fxl,Geng:2021iyq,Dong:2013qoa} for some related works. 
There is also double holography with higher-codimension branes; the typical examples include AdS/dCFT \cite{Jensen:2013lxa, DeWolfe:2001pq} and cone holography \cite{Miao:2021ual}. The branes with different codimensions present distinct features. For instance, the HM surface can be defined only in a finite time on codim-2 branes \cite{Hu:2022zgy}. Fortunately, it does not affect the Page curve since it happens after Page time for Einstein's gravity. 

In this paper, we generalize the discussions of \cite{Hu:2022zgy} to Gauss-Bonnet (GB) gravity. 
%GB gravity is a natural generalization of Einstein's gravity in higher dimensions, which includes only second-order derivatives. 
GB gravity represents a natural extension of Einstein's gravity into higher dimensions, primarily characterized by second-order derivatives. Besides, string theory predicts GB gravity as minor corrections of the gravitational action. 
%Furthermore, GB plays an essential role in AdS/CFT \cite{Buchel:2009sk}.
The significance of GB gravity is further underscored by its essential role in the AdS/CFT correspondence \cite{Buchel:2009sk} This paper investigates the black hole information problem in GB gravity with codim-2 branes. We verify that the Page curves can be recovered for all GB couplings within the causal constraints. Our results strongly support the island mechanism's universality and the causal constraints' plausibility. 

The paper is organized as follows. Section 2 reviews the AdS/dCFT correspondence for Gauss-Bonnet gravity. 
 Section 3 investigates Page curves on tensionless codim-2 brane in general dimensions. Additionally, we analyze GB parameters' effects on Page curves' behaviors. In section 4, we generalize the discussions to tensive codim-2 branes. Finally, we conclude with some discussions in section 5.

\section{Review of Gauss-Bonnet gravity and AdS/dCFT}

\begin{figure}[H]  %使得图片紧跟文字
	\begin{center}
		\begin{tikzpicture}
	\draw (0,0) node[below,scale=1.3]{$\text{D}$};
	\draw[color=red!80!black,line width=1.5pt] (0,0)--(6,0)node[midway,above]{$ \textcolor{black}{\text{AdS boundary}} $};
	\draw (3,0) node[below,scale=1.3]{$ M $};
%	\draw (7,0.5) node[above] {\text{Radiation}};
	\draw[color=red!80!black,line width=1.5pt] (6,0)--(8,0)node[midway,above]{$ \textcolor{black}{\text{}} $} ;
%	\draw (7,0) node[below,scale=1.3]{$ R $};
	\draw (0,8) node[above,scale=1.3]{$ r=0 $};
	\draw[color=black!85!white,line width=1.5pt] (0,0)--(0,6) node[midway,right]{$ \textcolor{black}{\text{Brane}} $} ;
	\draw (0,3) node[left,scale=1.3]{$ E $};
	\draw [->,line width=1.15pt] (-1.13,3) to node[left,scale=1.3]{$ z $} (-1.13,4);
	\draw[color=black!85!white,line width=1.5pt] (0,6)--(0,8);
%	\draw (0,7)  node[left,scale=1.3]{$ I $};
	\draw (8,0)node[right,scale=1.3]{$ r \rightarrow \infty  $};
	\draw[color=blue!90!yellow,line width=1.3pt] (6,0) arc(0:90:6) node[pos=0.5,above,sloped]{$\textcolor{black}{\text{Island RT Surface}}$};
	\draw[color=black,dashed,line width=1.5pt] (8,0) arc(0:90:8) node[midway,above,sloped]{$\text{Horizon}$};
	\draw [color=orange,line width=1.3pt](6,0)to [out=90,in=-80] node [pos=0.52,sloped,above]{$\textcolor{black}{\text{No-island RT Surface}}$} (5.657,5.657);
	\draw [->,line width=1.1pt] (5.7,6.85) arc (50:40:8.1) node[above,pos=0.6,scale=1.3]{$ r $};
	\filldraw (0,0) circle (0.05);
    \end{tikzpicture}
	\end{center}   %,bb=400 100 900 200
	\caption{The geometry of AdS/dCFT in black hole information paradox. For simplicity, we focus on the constant angle $\theta=0$ and the beginning time $t=0$.  The black line $E$ represents the codim-2 brane at $r=r_H$, where the hyperbolic black hole locates in. The CFT bath lives on the AdS boundary $M$ at $r\to \infty$.  The island surface and HM surface are labelled by blue and orange line separately. At early time, the HM surface has smaller area and is dominate. The area of HM surface increase over time while the area of island surface remains constant. As a result, the island surface is dominate at late times. }       %对图进行说明 
    \label{sketch}	
\end{figure}
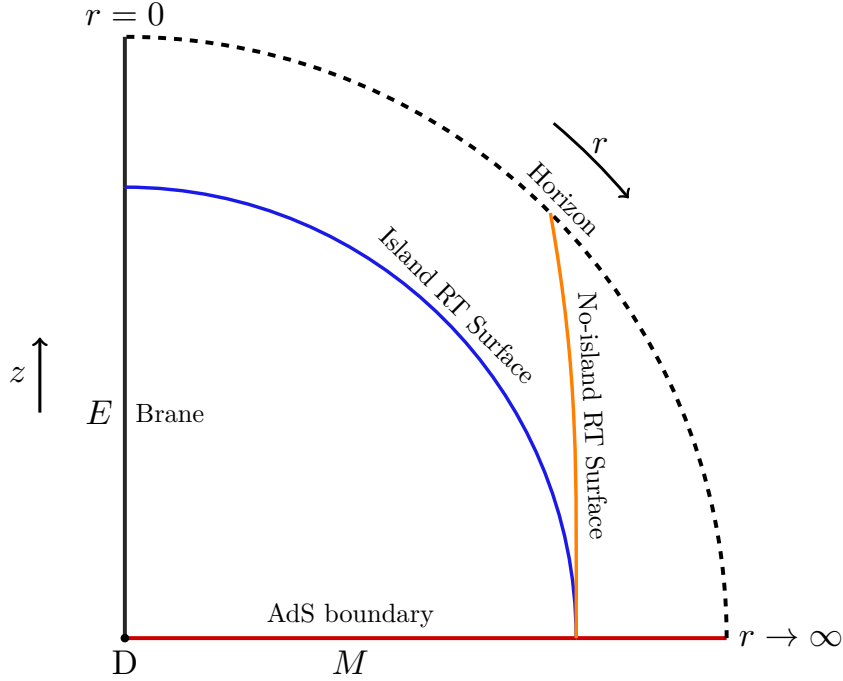

In this section, we give a brief review of AdS/dCFT for Gauss-Bonnet gravity.  Let us start with the geometry as shown in Fig.\ref{sketch}, where $M$ represents the AdS boundary and $E$ is the codim-2 brane, $D=\partial E$  is the codim-2 defect on AdS boundary $M$ \footnote{ Note that the defect $D$ is codim-2 on the AdS boundary and is codim-3 in bulk.}.  AdS/dCFT proposes that the classical gravity coupled with a codim-2 brane $E$ in bulk is dual to the CFT coupled with a codim-2 defect $D$ on the AdS boundary $M$. In the setup of brane world or double holography, our world with dynamical gravity (black hole) is defined on the brane $E$, and the CFT bath lives on the AdS boundary $M$. The entanglement entropy of Hawking radiation can be calculated by RT surfaces in bulk.  There are two kinds of RT surfaces, the island surface and the HM surface. The island surface (blue curve of Fig.\ref{sketch}) is perpendicular to both the brane and the AdS boundary. While the HM surface is perpendicular to the horizon at the beginning time $t=0$, then passes through the horizon as time evolved. The HM surface and island surface is dominate at early and late times, respectively.

The action of GB gravity is given by \cite{Myers:1987yn,Buchel:2009sk} 
\begin{align}
I_{GB}&=\frac{1}{16\pi G_N}\int_{N}^{}d^{d+1}x\sqrt{|g|}(R+\frac{d(d-1)}{L^2}+\frac{L^2\lambda_{GB}}{(d-2)(d-3)} (R_{\mu\nu\alpha\beta}R^{\mu\nu\alpha\beta}-4R_{\mu\nu}R^{\mu\nu}+R^2))\nonumber \\
&+\frac{1}{8\pi G_N}\int_{Q}^{}d^dy\sqrt{|h|}(K-T+\frac{2L^2\lambda_{GB}}{(d-2)(d-3)}(J-2G^{ij}_QK_{ij})).
\end{align}
Here we adopt the format outlined in \cite{Hu:2022ymx}, which just undergoes a reparameterization:
\begin{eqnarray}\label{action}
			I_{GB}=\int_N \dd[d+1]{x}\sqrt{\abs{g}}\pqty{R+\frac{d(d+1)}{l^2}+\alpha l^2(\bar{R}_{\mu\nu\alpha\beta}\bar{R}^{\mu\nu\alpha\beta}-4\bar{R}_{\mu\nu}R^{\mu\nu}+\bar{R}^2)}- T_E \int_E \sqrt{|h|},
\end{eqnarray}  
where we have set the Newton's constant to be $\frac{1}{16\pi\bar{G}_N}=1$, $R$ is the Ricci scalar, $l$ is the AdS radius, $T_E$ is the tension of codim-2 brane $E$, $\alpha$ is the GB coupling obeying the constraint. Following the reparameterization in \cite{Hu:2022ymx}, we have
\begin{eqnarray}
-\frac{1}{4\pqty{d^2-2d-2}}\leq \alpha \leq \frac{1}{8} \label{bcalpha},
\end{eqnarray}
and $\bar{R}$ is defined by \cite{Miao:2013nfa}
\begin{eqnarray}\label{barR}
&&R=\bar{R}-\frac{d(d+1)}{l^2},\\
		&&R_{\mu\nu}=\bar{R}_{\mu\nu}-\frac{d}{l^2}g_{\mu\nu},\\
		&&R_{\mu\nu\rho\sigma}=\bar{R}_{\mu\nu\rho\sigma}-\frac{1}{l^2}\pqty{g_{\mu\rho}g_{\nu\sigma}-g_{\mu\sigma}g_{\nu\rho}},
\end{eqnarray}
The bulk metric of AdS/dCFT is given by
     \begin{align}
\dd{s}^2_{\text{bulk}}=\frac{\dd{r}^2}{f(r)}+f(r)d\theta^2+r^2 \frac{\frac{\dd{z}^2}{1-z^2}-(1-z^2)\dd{t}^2+\dd{H}^2_{d-3}}{z^2},\label{frgeneral} 
\end{align}
where $\dd{H}^2_{d-3}=dy^2+\cosh^2(y)\dd{H}^2_{d-4}$ is the line element for a hyperbolic space, $f(r)=r^2-1$ for tensionless brane $E$, and $f(r)$ is given by:  
 \begin{align}
f(r)=&\frac{r^2\pqty{1+2\alpha(2-3d+d^2)}(1-\sqrt{1+M})}{2(d-2)(d-3)\alpha}-1 ,  \label{general f}   \\
M=&4\alpha r^{-d}(d-3)(d-2)\left[ -r^d r_H^4 \pqty{1+\alpha(d^2-d-2)}+r_H^d(r_H^2-1)\left(r_H^2\pqty{1+\alpha(d^2-d-2)} \right.\right. \nonumber\\
        &\left.\left. -\alpha(d^2-5d+6)\right)\right ]/\bqty{r_H^4 \pqty{1+2\alpha(d^2-3d+2)}^2},\label{M}
    \end{align}
 for tensive case, where we have reparameterized the metric of   \cite{Hung:2011nu}. Note that the brane $E$  locates at $r=r_H$ with $f(r_H)=0$. 
The holographic entanglement entropy for Gauss-Bonnet gravity in our context is originally given by \cite{Hung:2011xb}. Here after the preparameterization, it is now given by \cite{Hu:2022ymx}
\begin{align}\label{entropy}
    S_{\text{HEE}}=&4\pi\int_\Gamma \dd[d-1]{x}\sqrt{\gamma}\pqty{(1+2\alpha(d-1)(d-2))+2\alpha R_\gamma}+4\pi \int_{\partial \Gamma}\dd[d-2]{x}\sqrt{\sigma}4\alpha K_{\partial \Gamma}.
\end{align}
where $\Gamma$, $\gamma$ and $R_{\gamma}$ denote the RT surface, the determinant of induced metric on RT surface and Ricci scalar on the RT surface, separately. $\partial\Gamma$ is the boundary of RT surface where it intersects the brane. We use $\sigma$ to represent the determinant of induced metric on $\partial\Gamma$, and $K_{\partial\Gamma}$ is just the extrinsic curvature on $\partial\Gamma$.

\section{Page curve on tensionless codim-2 brane }

 In this section, we study the island and Page curve on the tensionless brane.  For simplicity, we focus on the eternal two-sided black hole, which is dual to the thermofield double state of CFTs \cite{Maldacena:2001kr}
\begin{eqnarray}\label{TFD}
| \text{TFD}\rangle=Z^{-1/2} \sum_{\alpha} e^{-E_{\alpha}/(2T)}e^{-iE_{\alpha}(t_L+t_R)} | E_{\alpha}\rangle_L  | E_{\alpha}\rangle_R,
\end{eqnarray}
where $L$ and $R$ label the states (times) associated with the left and right boundaries.  We derive general formulas for  $ \text{AdS}_{d+1}\text{/dCFT}_d $ but draw the Page curves only for $ \text{AdS}_6\text{/dCFT}_5$. Recall that the codim-2 brane $E$ is four dimensional in $ \text{AdS}_6\text{/dCFT}_5$, thus it is the most interesting case. We concentrate on hyperbolic black holes on the brane, where the bulk metric is  given by (\ref{frgeneral}) with $f(r)=r^2-1$. 

To set the geometric background for our analysis, we follow the approach outlined in \cite{Hu:2022zgy}. We draw the sketch in Fig \ref{sketch}, using polar coordinates (z, $\phi(r)$) with a constant $\theta$.

	\subsection{Island phase}\label{d=5 island}      
 \par First we consider the island phase, where RT surface, also called island surface, stays outside the horizon and does not evolve. See blue curve of Fig. \ref{sketch}. Assuming the embedding functions 
    \begin{equation}\label{ialand embed}
			t=\text{constant},\quad
			z=z(r).
	\end{equation}
Here we just study at some constant time t (since for eternal black hole, the island RT surface does not evolve over time), so the island RT surface can be determined by the function z(r). This allows us to %we 
derive the induced metric of the island surface:
\begin{align}\label{metric island}
\dd{s}^2_{\text{island}}=(\frac{1}{r^2-1}+&\frac{r^2z'(r)^2}{z(r)^2(1-z(r)^2)})\dd r^2+(r^2-1)\dd\theta^2+\frac{r^2}{z(r)^2}\dd H_{d-3}^2.
\end{align}
Now we want to obtain the entanglement entropy by substituting the above metric into the entropic formula (\ref{entropy}).%, we obtain the area functional 
The extrinsic curvature can be easily calculated by using:
\begin{align}
        K&=g^{\mu\nu} K_{\mu\nu}=g^{\mu\nu}(P^\alpha_\mu P^\beta_\nu\nabla_{(\alpha}n_{\beta)}),\\ 
        P_{\mu\nu}&=g_{\mu\nu}-n_{\mu}n_{\nu},
 \end{align}
which gives us a result: $\sqrt{\sigma} K_{\partial \Gamma}=-1/z(1)^{d-3}$ on the codim-2 brane $E$ located at $r=1$. Subsequently, R can be calculated by using the metric in \eqref{metric island}, and it is expressed as:
\begin{align}
    R_\gamma&=\frac{1}{2 r^2 f(r)^2 z(r)^2 \left(\frac{1}{f(r)}+\frac{r^2 z'(r)^2}{z(r)^2 \left(1-z(r)^2\right)}\right)^2}\left(r f(r) z(r) \left(2 (d-3) f(r) \left(z(r)-r z'(r)\right)+r z(r) f'(r)\right)\right. \nonumber\\
    &\left.\left(-\frac{f'(r)}{f(r)^2}-\frac{2 r^2 z'(r)^3}{z(r)^3 \left(1-z(r)^2\right)}+\frac{2 r^2 z'(r)^3}{z(r) \left(1-z(r)^2\right)^2}+\frac{2 r^2 z'(r) z''(r)}{z(r)^2 \left(1-z(r)^2\right)}+\frac{2 r z'(r)^2}{z(r)^2 \left(1-z(r)^2\right)}\right)\right.\nonumber\\
    &\left.+\left(\frac{1}{f(r)}+\frac{r^2 z'(r)^2}{z(r)^2 \left(1-z(r)^2\right)}\right) \left(-2 r f(r) z(r) \left(z(r) \left((d-3) f'(r)+r f''(r)\right)-(d-3) r f'(r) z'(r)\right)\right.\right.\nonumber\\
    &\left.\left.-2 (d-3) f(r)^2 \left(d r^2 z'(r)^2-2 r z(r) \left((d-2) z'(r)+r z''(r)\right)+(d-4) z(r)^2\right)+r^2 z(r)^2 f'(r)^2\right)\right.\nonumber\\
    &\left.-2 \left((d-1)^2-5 (d-1)+6\right) f(r)^2 z(r)^4 \left(\frac{1}{f(r)}+\frac{r^2 z'(r)^2}{z(r)^2 \left(1-z(r)^2\right)}\right)^2 \right).
\end{align}
For simplicity, we set the volume of hyperbolic space $H_{d-3}$ to be one. Collecting all the results above, we finally obtain the area functional: 
\begin{align}\label{entropyIsland}
 A_{\text{island}}&=\frac{S}{4\pi}=\int_\Gamma dr\left(\frac{r}{z(r)}\right)^{d-3} \sqrt{\frac{r^2 f(r) z'(r)^2}{z(r)^2-z(r)^4}+1} \left(2 \alpha  (d-2) (d-1)\right.\nonumber\\
    &\left.-\left(\alpha  \left(2 (d-3) f(r) \left(z(r)-r z'(r)\right)+r z(r) f'(r)\right) \left(z(r)^3 \left(f'(r)+2 r f(r)^2 z'(r) \left(r z''(r)+z'(r)\right)\right)+z(r)^7 f'(r)\right.\right.\right.\nonumber\\
    &\left.\left.\left.-2 z(r)^5 f'(r)-4 r^2 f(r)^2 z(r)^2 z'(r)^3+2 r^2 f(r)^2 z'(r)^3-2 r f(r)^2 z(r) z'(r) \left(rz''(r)+z'(r)\right)\right)\right)\right.\nonumber\\
    &\left./\left(r f(r) \left(r^2 f(r) z'(r)^2-z(r)^4+z(r)^2\right)^2\right)\right.\nonumber\\
    &\left.+\left(\alpha  \left(z(r)^2-1\right) \left(2 r f(r) z(r) \left((d-3) f'(r) \left(z(r)-r z'(r)\right)+r z(r) f''(r)\right)+2 (d-3) f(r)^2 \left(\left(z(r)-r z'(r)\right)\right.\right.\right.\right. \nonumber\\
    &\left.\left.\left.\left.\left((d-4) z(r)-d r z'(r)\right)-2 r^2 z(r) z''(r)\right)-r^2 z(r)^2 f'(r)^2\right)\right)/\left(r^2 f(r) \left(r^2 f(r) z'(r)^2-z(r)^4+z(r)^2\right)\right)\right.\nonumber\\
    &\left.-\frac{2 \alpha  (d-4) (d-3) z(r)^2}{r^2}+1\right)+ %\int_{\partial \Gamma}
    (\frac{1}{z(1)})^{d-3}4\alpha(-1).
\end{align}
%where we have used $\sqrt{\sigma} K_{\partial \Gamma}=-1/z(1)^{d-3}$ on the codim-2 brane $E$ located at $r=1$. For simplicity, we set the volume of hyperbolic space $H_{d-3}$ to be the one. 
Taking the variation of (\ref{entropyIsland}), we derive equations of motion (EOM) (see \eqref{eomr} in appendix (B) for details). From variation, the EOM we obtained just depicts the behaviour that the RT surface embedding function z(r) should follows, corresponding to the extreme value of the entanglement entropy. It's worth noting that the EOM \eqref{eomr} works for general $f(r)$. For the purposes of this section, we narrow our focus to the tensionless case with $f(r)=r^2-1$. 

%{\color{red}Next we need to obtain the boundary conditions (BCs) to completely solve z(r).}
Solving EOM \eqref{eomr} perturbatively around the brane $r=1$, we get BCs:
\begin{align} \label{BCisland}
	z(r)=z_{\text{brane}}+a_1(r-1)+O((r-1)^2),
\end{align}
with 
\begin{align}
	a_1&=\left(z_\text{brane} \left(\alpha  \left(-6 d^2+6 (d-4) (d-3) z_\text{brane}^2+38 d-64\right)-1\right)+\left(z_\text{brane}^2 \left(4 \alpha ^2 (d-2) \left(3 d^3\right.\right.\right.\right.\nonumber\\
    &\left.\left.\left.\left.-30 d^2+3 (d-4) (d-3)^2 z_\text{brane}^4-6 (d-5) (d-3) (d-2) z_\text{brane}^2+91 d-80\right) \right.\right.\right.\nonumber\\
    &\left.\left.\left.+4 \alpha  \left(3 (d-3) z_\text{brane}^2-d+5\right)+1\right)\right)^\frac12\right)/(12\alpha  (d-3)) ,
\end{align}
and $z_{\text{brane}}$ denotes the value of $z(r)$ on the brane $E$. Because of the complexity of EOM, it is almost impossible to get an analytical solution. Consequently, we resort to numerical calculations. %{\color{red}Here we imposed the boundary conditions above for the sake of numerical calculation.}%
By applying the shooting method, we can obtain the island surface $z(r)$ by numerically solving EOM (\ref{eomr}) with BCs (\ref{BCisland}) on the codim-2 brane $E$ and adjusting $z_\text{zbrane}$ to fit BC: $z(r_{UV})=z_{\text{bdy}}$, with any given $z_\text{bdy}$ on the AdS boundary $M$. Here $z_{\text{bdy}}$ denotes the value of $z(r)$ on the AdS boundary $r=r_\text{UV}$, with $r_\text{UV}$ serves a cutoff.

\subsection{No-island phase}\label{d=5 noisland}
	
\begin{figure}[t]  
	\centering
	\includegraphics[width=0.6\textwidth]{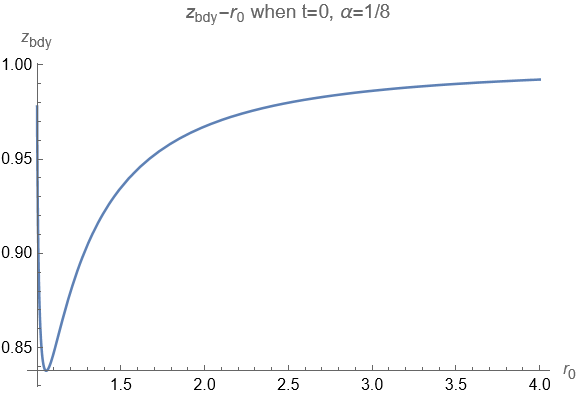}
	\caption{The relation between $z_{\text{bdy}}$ and $r_0$ for $\alpha=\frac18$. It shows that there is a minimum value of $z_{\text{bdy}}$ in the no-island phase.} 
	\label{d=5 zbdy-r0} 
\end{figure}

Let us go on to study the the RT surface in the no-island phase, which is named the HM surface. At the beginning time $t_R=t_L=0$, the HM surface is perpendicular to both the horizon and AdS boundary. Solving EOM (\ref{eomr}) near the horizon $z=1$, we get
% \begin{align}
% 	z(r)&=1+b_1(r-r_0)+b_2(r-r_0)^2+O((r-r_0)^3)\\
% 	b_1&=\frac{-r_0^2+12\alpha-48r_0^2\alpha}{r_0(-3r_0^2+4r_0^4+12\alpha-96r_0^2\alpha+96r_0^4\alpha)}
% \end{align}
\begin{align}
    z(r)&=1+b_1(r-r_0)+O((r-r_0)^2),
\end{align}
where $b_1$ is
\begin{align}
    &\frac{6 \alpha  \left(d^2-7 d+12\right)-r_0^2 \left(\alpha  \left(6 d^2-26 d+28\right)+1\right)}{(d-1) r_0^5 \left(2 \alpha  \left(d^2-3 d+2\right)+1\right)-(d-2) r_0^3 \left(2 \alpha  \left(2 d^2-9 d+11\right)+1\right)+2 \alpha  \left(d^3-9 d^2+26 d-24\right) r_0}.
\end{align}
For any given $r_0$, one can numerically obtain $z(r)$ by solving EOM (\ref{eomr}) with the above BCs. Then, one can derive $z_{\text{bdy}}=z(r_{UV})$. Just like the case in Einstein gravity, there is a lower bound for $z_{\text{bdy}}$.  See Fig \eqref{d=5 zbdy-r0} for an example with $\alpha=1/8$. According to \cite{Hu:2022zgy}, the lower bound of $z_{\text{bdy}}$ leads to an upper bound for the time $t_R=t_L$. 

As time evolves, the HM surface crosses the horizon. To avoid coordinate singularity, we choose the Eddington-Finkelstein coordinate $ v=t-\frac{1}{2}\log(\abs{\frac{1+z}{1-z}}) $. Then the time and metric become
    \begin{equation}\label{boundary time}
		t_R=v(z_{\text{bdy}})+\frac{1}{2}\log(\abs{\frac{1+z_{\text{bdy}}}{1-z_{\text{bdy}}}}),
    \end{equation}
and
 \begin{equation}
	\dd{s^2}=\frac{\dd{r}^2}{r^2-1}+(r^2-1)\dd{\theta}^2+r^2\frac{-(1-z^2)\dd{v^2}-2\dd{v}\dd{z}+\dd{H}^2_{d-3}}{z^2}.
\end{equation}
Assuming the embedding functions $r=r(z),\ v=v(z)$, we get the induced metric for HM surface
\begin{align}\label{HMmetric}
\dd{s^2}=&\pqty{\frac{r'(z)^2}{r(z)^2-1}-\frac{r(z)^2(1-z^2)v'(z)^2}{z^2}-\frac{2r(z)^2v'(z)}{z^2}}\dd{z^2}+(r(z)^2-1)\dd{\theta}^2+\frac{r(z)^2}{z^2}\dd{H}^2_{d-3}.
\end{align}
From (\ref{entropy}) and (\ref{HMmetric}), we obtain the area functional for HM surface
\begin{align}\label{entropyHM}
    A_{\text{HM}}&=\frac{S}{4\pi}=\int_\Gamma dz \frac{r(z)^{d-3}}{z^{d-3}} \sqrt{\frac{r(z)^2 f(r(z)) v'(z) \left(\left(z^2-1\right) v'(z)-2\right)}{z^2}+r'(z)^2}
   %  \left(\prod _{j=1}^{d-3} \cosh ^{d-j-3}(y(j))\right)
     \nonumber\\
    &\left(2 \alpha  (d-2) (d-1)+\frac{\alpha  z^2}{r(z)^2} \left(\left(r(z) \left(z r(z) r'(z) f'(r(z))-2 (d-3) f(r(z)) \left(r(z)-z r'(z)\right)\right) \left(2 f(r(z))\right.\right.\right.\right.\nonumber\\
    &\left.\left.\left.\left.\left(r(z) f(r(z)) \left(v'(z) \left(z r'(z) \left(\left(z^2-1\right) v'(z)-2\right)+r(z) \left(v'(z)+2\right)\right)+z r(z) \left(\left(z^2-1\right) v'(z)-1\right) v''(z)\right)\right.\right.\right.\right.\right.\nonumber\\
    &\left.\left.\left.\left.\left.+z^3 r'(z) r''(z)\right)-z^3 r'(z)^3 f'(r(z))\right)\right)/\left(f(r(z)) \left(z r(z)^2 f(r(z)) v'(z) \left(\left(z^2-1\right) v'(z)-2\right)+z^3 r'(z)^2\right)^2\right)\right.\right. \nonumber\\
    &\left.\left.+\left(-2 z r(z) f(r(z)) \left(r'(z) \left((d-3) \left(z r'(z)-r(z)\right) f'(r(z))+z r(z) r'(z) f''(r(z))\right)+z r(z) r''(z) f'(r(z))\right)\right.\right.\right.\nonumber\\
    &\left.\left.\left.-2 (d-3) f(r(z))^2 \left(\left(r(z)-z r'(z)\right) \left(d r(z)-(d-4) z r'(z)\right)+2 z^2 r(z) r''(z)\right)+z^2 r(z)^2 r'(z)^2 f'(r(z))^2\right)\right.\right.\nonumber\\
    &\left.\left./\left(z^2 f(r(z)) \left(r(z)^2 f(r(z)) v'(z) \left(\left(z^2-1\right) v'(z)-2\right)+z^2 r'(z)^2\right)\right)-2 (d-4) (d-3)\right)+1\right).
\end{align}
Note that there is no boundary term $\sqrt{\sigma} K_{\partial \Gamma}$ in the no-island phase,  since the HM surface does not intersect the brane $E$. Taking the variation of (\ref{entropyHM}), we can derive the EOM. See \eqref{EOMHM} in appendix (B) for details. 
\par Following \cite{Carmi:2017jqz,Hu:2022zgy}, we impose the following BCs on the turning point $z=z_{\text{max}}$ and $r=r_0$ for the two-side black holes
\begin{align}\label{bc5}
	v(z_{\text{max}})&=-\frac12\log(\frac{z_{\text{max}}+1}{z_{\text{max}}-1}),\  v'(z_{\text{max}})=-\infty, \nonumber\\
	r(z_{\text{max}})&=r_0 ,\quad  r'(z_{\text{max}})= g(r_0,\alpha,z_{\text{max}}).
\end{align}
The first line represents the boundary conditions from reference \cite{Carmi:2017jqz}, obtained through the conditions $t(z_\text{max}) = 0$ and the symmetry of turning point. In the second line, $g(r_0,\alpha,z_{\text{max}})$ is a complicated equation fixed by EOM. We obtain:
\begin{align}\label{g5dHM}
 g(r_0,\alpha,z_{\text{max}})&=\left(r_0 f(r_0) \left(r_0 f'(r_0) \left(r_0^4 \left(2 \alpha  \left(d^2-3 d+2\right)+1\right)^2-4 \alpha  (d-3) r_0^2 \left(2 \alpha  \left(d^2-3 d+2\right)+1\right) \right.\right.\right.\nonumber\\
 &\left.\left.\left.\left((d-5)z_{\text{max}}^2+2\right)+4 \alpha ^2 (d-3)^2 \left(d^2-6 d+8\right) z_{\text{max}}^4\right)+2 (d-2) f(r_0)\left(r_0^2 \left(2 \alpha \right.\right. \right.\right.\nonumber\\
&\left.\left.\left.\left.\left(d^2-3 d+2\right)+1\right)-2 \alpha  \left(d^2-7 d+12\right) z_{\text{max}}^2\right)^2\right)\right)/\left(2 z_{\text{max}} \left(-\alpha  (d-3) r_0^2 \left(z_{\text{max}}^2-1\right)\right.\right.\nonumber\\
&\left.\left.f'(r_0)^2 \left(r_0^2 \left(2 \alpha  \left(d^2-3 d+2\right)+1\right)-2 \alpha  \left(d^2-7 d+12\right) z_{\text{max}}^2\right)+f(r_0) \left(-4 \alpha  (d-3) r_0 f'(r_0)\right.\right.\right.\nonumber\\
&\left.\left.\left.\left(r_0^2 \left(2 \alpha  \left(d^2-3 d+2\right)+1\right)-2 \alpha  \left(d^2-7 d+12\right) z_{\text{max}}^4\right)+r_0^4 \left(2 \alpha  \left(d^2-3 d+2\right)+1\right)^2\right.\right.\right.\nonumber\\ 
&\left.\left.\left.\left(d \left(z_{\text{max}}^2-1\right)-3 z_{\text{max}}^2+2\right)-4 \alpha  \left(d^2-7 d+12\right) r_0^2 z_{\text{max}}^2 \left(2 \alpha  \left(d^2-3 d+2\right)+1\right)\right.\right.\right. \nonumber\\
&\left.\left.\left.\left(d \left(z_{\text{max}}^2-1\right)-4 z_{\text{max}}^2+3\right)+4 \alpha ^2 \left(d^2-7 d+12\right)^2 z_{\text{max}}^4 \left(d \left(z_{\text{max}}^2-1\right)-5 z_{\text{max}}^2+4\right)\right)\right.\right.\nonumber\\
&\left.\left.-4 \alpha  \left(d^2-7 d+12\right) z_{\text{max}}^2 f(r_0)^2 \left(r_0^2 \left(2 \alpha  \left(d^2-3 d+2\right)+1\right)-2 \alpha  \left(d^2-7 d+12\right) z_\text{max}^2\right)\right)\right).
\end{align}
\begin{figure}[t]
    \centering
    \includegraphics[width=10cm]{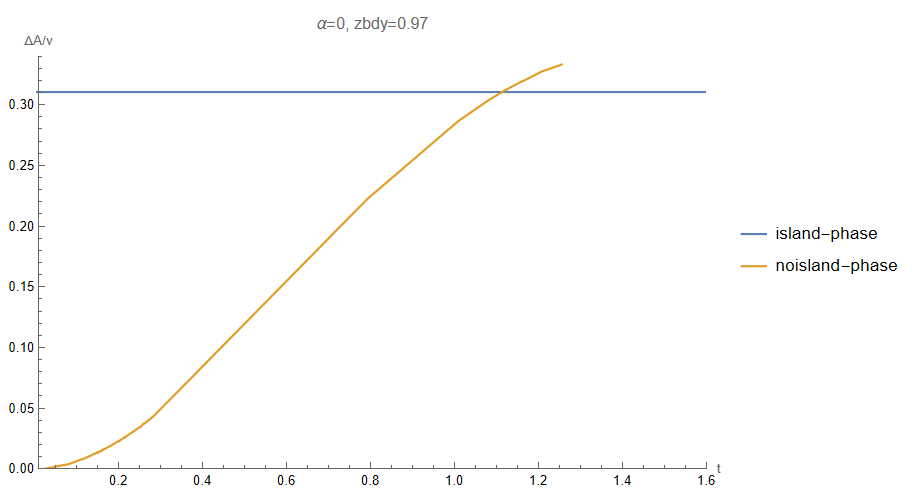}
    \caption{The Page curve for Einstein gravity with $\alpha=0$, $d=5, z_{\text{bdy}}=0.97$. Here  $\Delta A=A(t)-A(0)$ denotes the area difference, $V=\int d\theta  dH_{d-3}$ denotes the angle period multiplied by the volume of hyperbolic space. The entanglement entropy of Hawking radiation first increases over time (orange curve), and then becomes a constant (blue curve), which recovers the Page curve of eternal black holes.}
    \label{d=5 a=0}
\end{figure}

% \begin{figure}[t]
%     \centering
%     {\includegraphics[width=7.3cm]{a18zbdy97.png}}
%     \quad
%     {\includegraphics[width=7.3cm]{an152zbdy99.png}}
%     \caption{The page curves when $\alpha=1/8, -1/52$. The figure on the left shows the case when $\alpha=1/8, z_{\text{bdy}}$=0.97, The figure on the right gives the result when $\alpha$=-1/52, $z_{\text{bdy}}$=0.99, where the curves of island phase and no-island phase could not intersect.  }
%     \label{d=5 tensionless a>0}
% \end{figure}

\begin{figure}[H]
    \centering
    {\includegraphics[width=7.3cm]{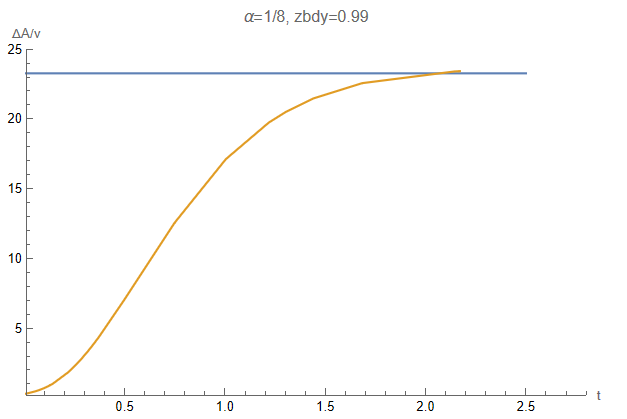}}
    \quad
    {\includegraphics[width=7.3cm]{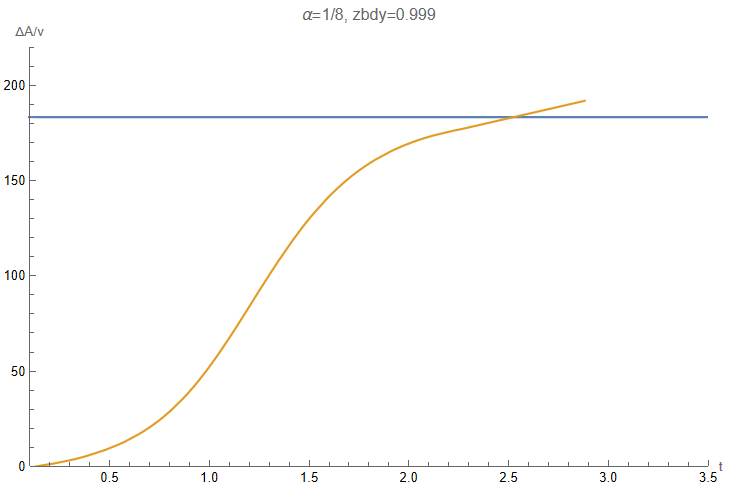}}
    {\includegraphics[width=7.3cm]{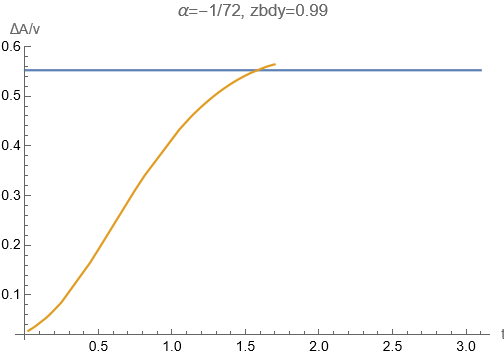}}
    \quad
    {\includegraphics[width=7.3cm]{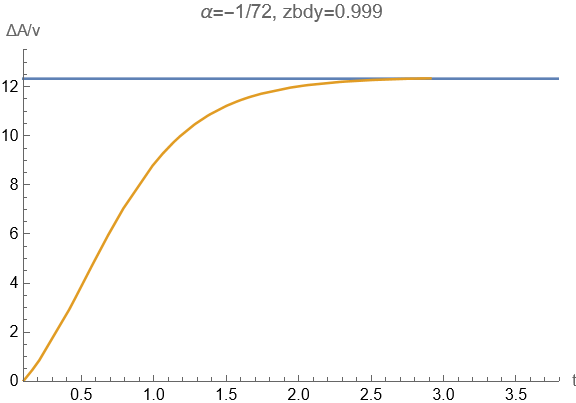}}
     {\includegraphics[width=7.3cm]{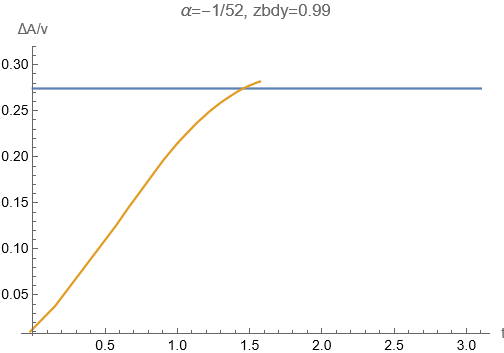}}
    \quad
    {\includegraphics[width=7.3cm]{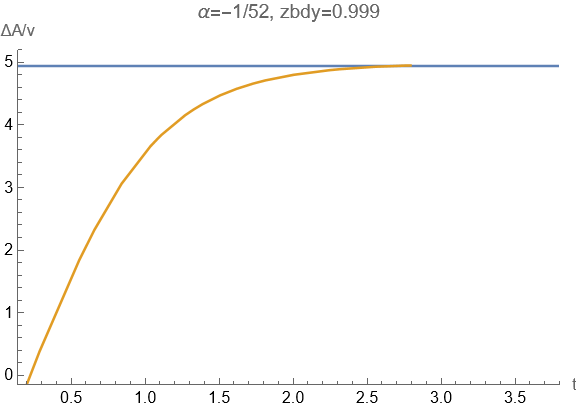}}
    \caption{Page curves for GB gravity with various couplings $\alpha$. Recall that $\Delta A=A(t)-A(0)$ and $V=\int d\theta dH_{d-3}$. The parameters for the upper, middle and lower figures are given by ($\alpha$=1/8,$z_{\text{bdy}}$=0.99, 0.999),($\alpha$=-1/72,$z_{\text{bdy}}$=0.99, 0.999) and ($\alpha$=-1/52, $z_{\text{bdy}}$=0.99, 0.999), respectively.  For fixed $\alpha$, the area difference $\Delta A=A(t)-A(0)$ and the Page time increase with $z_{\text{bdy}}$. And for fixed $z_{\text{bdy}}$, the area difference $\Delta A$  and Page time increase with $\alpha$.  } 
\label{d=5 tensionless}
\end{figure}

From EOM (\ref{EOMHM}) and BCs \eqref{bc5}, we can numerically solve functions $v(z)$ and $r(z)$ and then calculate the area of HM surface and boundary time via equations \eqref{entropyHM} and \eqref{boundary time}. It enables us to derive the Page curve. See Fig. \ref{d=5 a=0} for the case of Einstein gravity within $\alpha=0$ and Fig. \ref{d=5 tensionless} for GB gravity within causal constraints $\frac{-1}{52}\leq\alpha\le\frac18$. The Page curves are given by the orange curves at the early times and the blue curves after the Page time. We find that the Page curves of eternal black holes can be recovered for all of the GB couplings within the causal constraints. Our results are strong support for the universality of the island mechanism. 

To end this section, we calculate the Page time and maximum time of HM surface for various parameters $z_{\text{bdy}}$ and $\alpha$. We list the results in Table \ref{table1} and Table \ref{table2}. Table \ref{table1} shows that both times increase with $z_{\text{bdy}}$, and Table \ref{table2} shows that the Page time and the maximum time are positively correlated with $\alpha$. We also find that the entropy increased when increasing $\alpha$ or $z_{bdy}$. Like Einstein's gravity, we find that the HM surface can exist only in a finite time. Remarkably, for various parameters, $\alpha$ and $z_{\text{bdy}}$, the maximum times of HM surface are always larger than the Page times. Thus, this strange behavior of HM surfaces does not affect the Page curves for GB gravity.
% \begin{figure}[H]
%     \centering
%     {\includegraphics[width=7.3cm]{a18zbdy99.png}}
%     \quad
%     {\includegraphics[width=7.3cm]{a18zbdy999.png}}
% \end{figure}

\begin{table}[H]
\begin{minipage}{0.5\linewidth}
\centering
\caption{Time with different $z_{\text{bdy}}$ ($\alpha=1/8$)}\label{table1}
\begin{tabular}{c|c|c} 
\toprule
$z_{\text{bdy}}$ & Page time & Maximum time\\
\midrule
 0.95 &1.091 & 1.271 \\
 0.96 & 1.235 &1.432\\
0.97 &1.408 &1.598\\  
0.98 & 1.639& 1.825\\
0.99 &2.011 &2.170\\  
\bottomrule
\end{tabular} 
\end{minipage}\begin{minipage}{0.5\linewidth}  
\centering
\caption{Time with different $\alpha$ ($z_{\text{bdy}}=0.99$)} \label{table2}
\begin{tabular}{c|c|c} 
\toprule
$\alpha$ & Page time & Maximum time\\
\midrule
-1/300 & 1.718  &1.849\\
-1/1000& 1.740  &1.872\\
1/32 &   1.897 &2.050\\
1/16 & 1.957   &2.124\\
1/8 &  2.011  & 2.170\\
\bottomrule  
\end{tabular} 
\end{minipage}
\end{table}

% \begin{table}[h]
% \centering
% \caption{Time with different $z_{\text{bdy}}$ ($\alpha=1/8$)}\label{table1}
% \begin{tabular}{ccc}
% \toprule
% $z_{\text{bdy}}$ & Page time & Maximum time\\
% \midrule
%  0.95 &1.091 & 1.271 \\
%  0.96 & 1.235 &1.432\\
% 0.97 &1.408 &1.598\\  
% 0.98 & 1.639& 1.825\\
% 0.99 &2.011 &2.170\\  
% \bottomrule
% \end{tabular}
% \end{table}

% \begin{table}[h]
% \centering
% \caption{Time with different $\alpha$ ($z_{\text{bdy}}=0.99$)} \label{table2}
% \begin{tabular}{ccc}
% \toprule
% $\alpha$ & Page time & Maximum time\\
% \midrule
% -1/300 & 1.718  &1.849\\
% -1/1000& 1.740  &1.872\\
% 1/32 &   1.897 &2.050\\
% 1/16 & 1.957   &2.124\\
% 1/8 &  2.011  & 2.170\\
% \bottomrule  
% \end{tabular}
% \end{table}

\newpage
\section{Page curve on tensive codim-2 branes}

In this section, we go on to study Page curves on the brane with non-zero tension. The behavior of Page curves is similar to that of Einstein gravity \cite{Hu:2022zgy}. For instance, the maximum time of HM surface increases with the tension and tends to the Page time from above. Since the calculations are similar to tensionless case of sect. 3, we only show the main result below. The main difference is that,  instead of $f(r)=r^2-1$, now $f(r)$ is given by \eqref{general f}:
 \begin{align}
     f(r)=-1+\frac{r^2(1+24\alpha-\sqrt{(1+12\alpha)^2+\frac{24r_H(-1+r_H^2)\alpha(-6\alpha+r_H^2(1+18\alpha))}{r^5}})}{12\alpha},
 \end{align}
for $d=5$, where $f(r_H)=0$.
 
\begin{figure}[H]
        \centering
        \includegraphics[width=10cm]{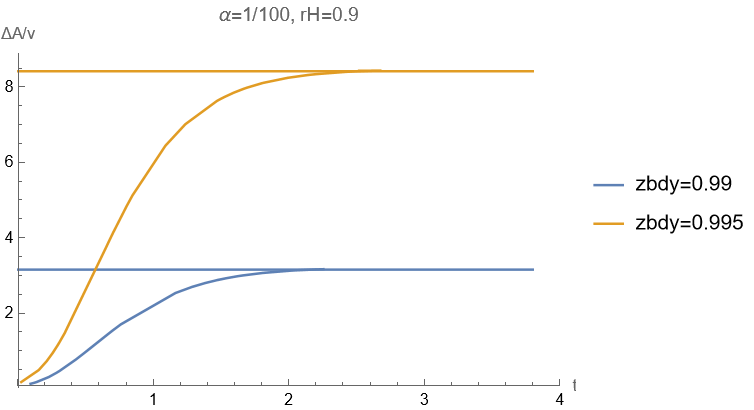}
        \caption{Tensive Page curve for $ \alpha=1/100 $ and $ r_{\text{H}}=0.9$. The blue and orange lines are for $ z_{\text{bdy}}=0.99 $ and $ z_{\text{bdy}}=0.995 $, respectively. It shows that the Page time increases with $z_{\text{bdy}}$.}
        \label{d=5 tensive1}
    \end{figure}

\begin{figure}[H]
    \centering
    {\includegraphics[width=7.5cm]{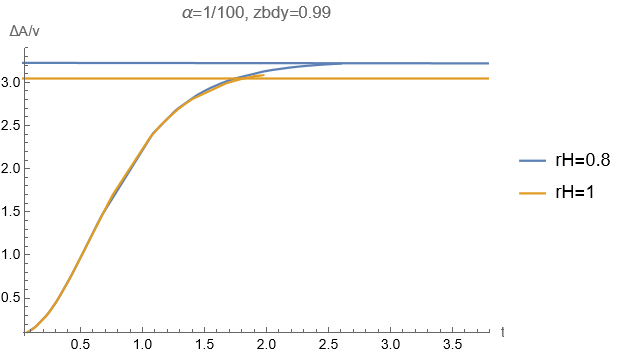}}
     {\includegraphics[width=7.5cm]{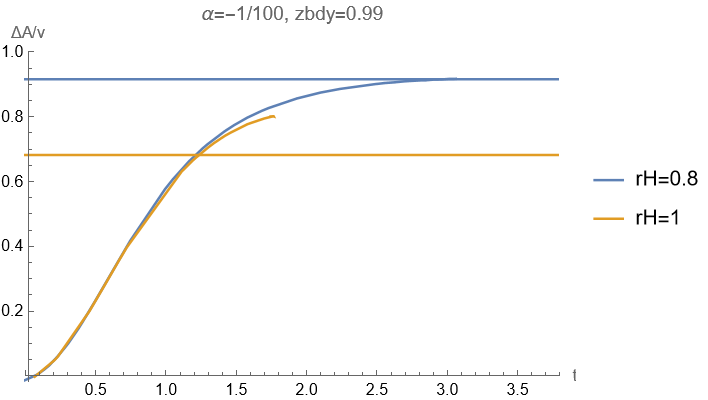}}
     {\includegraphics[width=7.5cm]{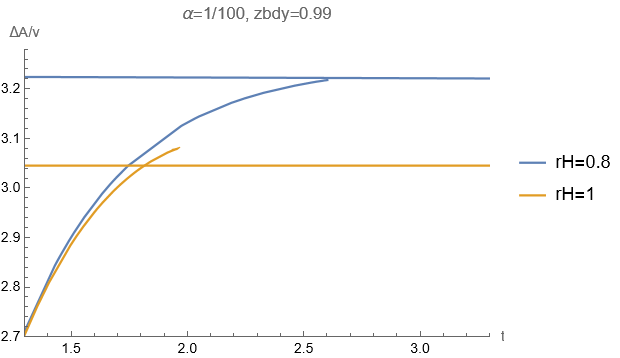}}
     {\includegraphics[width=7.5cm]{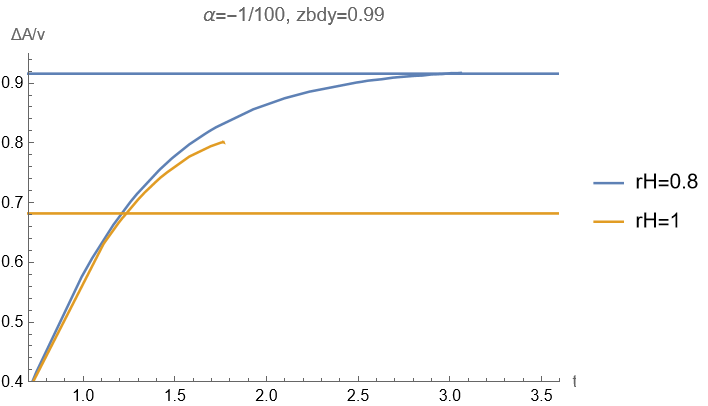}}
    \caption{(left)Tensive Page curves for $\alpha=1/100$,  \text{$r_H=1, 0.8$}, from bottom to top; (right) Tensive Page curves for $\alpha=-1/100$, \text{$r_H=1, 0.8$}, from bottom to top. Here we choose $z_{\text{bdy}}$=0.99. The figures below give the enlargement of the interested parts. It shows that the Page time decreases with $r_H$ (also means the Page time increases with the tension).}
    \label{d=5 tensive2}
\end{figure}

We plot the Page curve of tensive branes in Fig. \ref{d=5 tensive1} and Fig. \ref{d=5 tensive2} .  In Fig. \ref{d=5 tensive1}, we consider the Page curves with the same $\alpha, r_H$ but different  $z_{\text{bdy}}$. It shows that the Page time increases with $z_{\text{bdy}}$, which is similar to the tensionless case.  Note that $r_H \le 1$ decreases with the tension of the brane \cite{Hu:2022ymx}.  In Fig. \ref{d=5 tensive2}, we discuss the case with the same $\alpha$ and $z_{\text{bdy}}$ but different $r_H$. It turns out that the brane with higher tension (lower $r_H$) tends to have larger Page times.

\section{Conclusions and discussions }	
In summary, this paper investigates the black hole information paradox on the codim-2 brane for GB gravity. Due to the island surfaces ending on the codim-2 branes, the Page curves of eternal black holes can be recovered for a general range of GB couplings. We show the island rules works well for doubly holographic models with higher derivative gravity and codim-2 branes, which strongly support the universality of islands. 

Analogous to the case in Einstein gravity,  for no-island phase, the extremal surface cannot be defined after some finite time, but this does not affect the Page curve, since it happens after the Page time. 
%Moreover, we find that the Page times increase with the GB parameter $\alpha$ while decreasing with the tension of branes. This similar trend also happens on the maximum time. 
Furthermore, our investigation has revealed a clear correlation between the Page times and the Gauss-Bonnet parameter $\alpha$. Specifically, the Page time exhibits a discernible increase as $\alpha$ is augmented, and concurrently, the tension of the branes imparts a proportional effect on the Page times. This analogous pattern is also observed in the maximum time.

%In the future, it is interesting to generalize the discussions to other kinds of higher derivative gravity and evolving black holes.
In future research, it would be valuable to expand upon these discussions and consider other variations of higher derivative gravity and their implications for evolving black holes. Understanding these aspects could deepen our comprehension of the fundamental nature of gravity and its effects on black hole evolution and information dynamics.

%%%%% Citations in the text %%%%%%
\section*{Acknowledgements}
We be grateful for R. X. Miao for kind guidances and valuable discussions. We thank the supports from ``College Students' Innovative Entrepreneurial Training Plan Program". \\

\section*{Appendix A: Boundary conditions for tensive case }

In this appendix, we list the BCs for calculations of Page curves for tensive branes with $d=5$. 

{\bf BCs for Island phase}
	\begin{equation}
		z(r)=z_{\text{brane}}+c_1(r-r_H)+\order{r-r_H}^2.
	\end{equation}
        For d=5, we have:
        \begin{align}
            c_1&=\frac{-B+\sqrt{B^2-4AC}}{2A},\\            A&=\frac{12r_Hz_{\text{brane}}^3(-1+z_{\text{brane}}^2)^2\alpha(6\alpha-3r_H^2(1+24\alpha)+r_H^4(5+90\alpha))^2}{(-12\alpha+r_H^2(1+24\alpha))^2}\\
            B&=\left\{2z_{\text{brane}}^4(z_{\text{brane}}^2-1)^2(6\alpha-3r_H^2(1+24\alpha)+r_H^4(5+90\alpha))(24\alpha^2(-5+6z_{\text{brane}}^2)\right. ,\nonumber\\            &\left.-4r_H^2(-2+3z_{\text{brane}}^2)\alpha(1+24\alpha)+r_H^4(1+4\alpha(7+54\alpha)))  \right\}/(-12\alpha+r_H^2(1+24\alpha))^2,\\ 
            C&=-\left\{ 4z_{\text{brane}}^5(z_{\text{brane}}^2-1)^3\left(-846\alpha^4+288r_H^2\alpha^3(1+24\alpha)+24r_H^4\alpha^2(11+288\alpha+936\alpha^2)\right. \right.\notag\\ &\left.\left.-48r_H^6\alpha(1+47\alpha+678\alpha^2+3024\alpha^3)+r_H^8(1+96\alpha+2856\alpha^2+33696\alpha^3+137376\alpha^4)\right)      \right\}\notag\\
            &/\bqty{r_H\pqty{r_H^2(1+24\alpha)-12\alpha}^3}.
        \end{align}

{\bf BCs for no-Island phase at $t=0$}
	\begin{equation}
		z(r)=1+d_1(r-r_0)+\order{r-r_0}^2,
	\end{equation}
 where 
\begin{align}
     d_1&=-2(r_0^2-4\alpha+24\alpha r_0^2+8\alpha f(r_0)+8r_0\alpha f'(r_0))/ \left\{ r_0(6r_0^2f(r_0)-24\alpha f(r_0)\right. \nonumber\\
       &\left.+144r_0^2\alpha f(r_0)+r_0^3f'(r_0)-12r_0\alpha f'(r_0)+24r_0^3\alpha f'(r_0))\right\}.
\end{align}

{\bf BCs for no-Island phase at $t>0$}

The BCs are given by (\ref{bc5}) with
\begin{align}\label{gappendixHM}
    g(r_0,\alpha,z_{\text{max}})&=-\left(r_0 f(r_0) \left(r_0 f'(r_0) \left((24 \alpha +1)^2 r_0^4-16 \alpha  (24 \alpha +1) r_0^2+48 \alpha ^2 z_{\text{max}}^4\right)+6 f(r_0) \left((24 \alpha +1) r_0^2\right.\right.\right.\nonumber\\
    &\left.\left.\left.-4 \alpha  z_{\text{max}}^2\right)^2\right)\right)/\left(2 z_{\text{max}} \left(2 \alpha  r_0^2 \left(z_{\text{max}}^2-1\right) f'(r_0)^2 \left((24 \alpha +1) r_0^2-4 \alpha  z_{\text{max}}^2\right)+f(r_0)\right.\right.\nonumber\\
    &\left.\left.\left(8 \alpha  r_0 f'(r_0) \left((24 \alpha +1) r_0^2-4 \alpha  z_{\text{max}}^4\right)-\left((24 \alpha +1)^2 r_0^4 \left(2 z_{\text{max}}^2-3\right)\right)+8 \alpha  (24 \alpha +1) r_0^2 z_{\text{max}}^2\right.\right.\right.\nonumber\\
    &\left.\left.\left.\left(z_{\text{max}}^2-2\right)+16 \alpha ^2 z_{\text{max}}^4\right)+8 \alpha  z_{\text{max}}^2 f(r_0)^2 \left((24 \alpha +1) r_0^2-4 \alpha  z_{\text{max}}^2\right)\right)\right)
\end{align} 

\section*{Appendix B: Equations of motion for both Island and No-island phase}

In this appendix, we list the EOMs for calculations of Page curves for island and no-island phase in any dimension $d$. Notice that these EOMs are suitable for both tensionless and tensive cases by simply changing $f(r)$.

{\bf  EOMs for Island phase}
\begin{align}\label{eomr}
    &z''(r)=-\left(-4 (d-5) (d-4) (d-3) \alpha  z(r)^{13}+2 (d-3) \left((2 (d-2) (d-1) \alpha +1) r^2-2 \alpha  f''(r) r^2\right.\right.\nonumber\\
    &\left.\left.+6 (d-5) (d-4) \alpha -2 (d-4) \alpha  \left((d-5) f(r)+2 r f'(r)\right)\right) z(r)^{11}+12 (d-4) (d-3) r \alpha  \left((d-3) f(r)\right.\right.\nonumber\\
    &\left.\left.+r f'(r)\right) z'(r) z(r)^{10}+2 (d-3) \left(-3 (2 (d-2) (d-1) \alpha +1) r^2+6 \alpha  f''(r) r^2-6 (d-5) (d-4) \alpha\right.\right. \nonumber\\
    &\left.\left.+2 (d-4) \alpha  \left(6 r f'(r)+f(r) \left((d-11) r^2 z'(r)^2+3 (d-5)\right)\right)\right) z(r)^9-2 r z'(r) \left(-2 (d-4) (d-3)^2 \right.\right.\nonumber\\
    &\left.\left.\alpha  f(r)^2+\left(16 (d-4) \alpha  (d-3)^2-2 r \alpha  \left((3 d-10) f'(r)+r f''(r)\right) (d-3)+(d-1) r^2 (2 (d-2))\right.\right.\right.\nonumber\\
     &\left.\left.\left.(d-1) \alpha +1)\right) f(r)+r f'(r) \left(r^2-2 (d-3) \alpha  f'(r) r+2 \left((d-2)(d-1) r^2+8 (d-4) (d-3)\right) \alpha \right)\right)\right.\nonumber\\
     &\left.z(r)^8-2 \left(2 \left(d^3-37 d+84\right) r^2 \alpha  f(r)^2 z'(r)^2+(d-3) \left(-3 (2 (d-2) (d-1) \alpha +1) r^2+3 \alpha \right.\right.\right.\nonumber\\
     &\left.\left.\left.\left(f'(r)^2 z'(r)^2 r^3+2 f''(r) r+4 (d-4) f'(r)\right) r-2 (d-5) (d-4) \alpha \right)+2 f(r) \left(r^2 \left((d-4) \left((2 (d-2)\right.\right.\right.\right.\right.\nonumber\\
     &\left.\left.\left.\left.\left.(d-1) \alpha +1) r^2+(d-20) (d-3) \alpha \right)+(d-3) r \alpha  \left((3 d-5) f'(r)-r f''(r)\right)\right) z'(r)^2+3 (d-5)\right.\right.\right.\nonumber\\
     &\left.\left.\left.(d-4) (d-3) \alpha \right)\right) z(r)^7-2 r z'(r) \left(2 (d-4) (d-3) \alpha  \left(3 (d-8) r^2 z'(r)^2+2 (d-3)\right) f(r)^2\right.\right.\nonumber\\
     &\left.\left.+\left(-14 (d-4) \alpha  (d-3)^2+r \alpha  \left(f'(r) \left(3 (d-8) r^2 z'(r)^2+12 d-40\right)+4 r f''(r)\right) (d-3)-2 (d-1)\right.\right.\right.\nonumber\\
     &\left.\left.\left.r^2 (2 (d-2) (d-1) \alpha +1)\right) f(r)+2 r f'(r) \left(-\left((2 (d-2) (d-1) \alpha +1) r^2\right)+2 (d-3) \alpha  f'(r) r\right.\right.\right.\nonumber\\
     &\left.\left.\left.-7 (d-4) (d-3) \alpha \right)\right) z(r)^6+2 \left(2 (d-4) (d-3) (2 d+13) r^2 \alpha  f(r)^2 z'(r)^2-(d-3) r \left(2 (d-2) \right.\right.\right.\nonumber\\
     &\left.\left.\left.(d-1) \alpha  r-2 \alpha  f''(r) r+r-2 \alpha  f'(r) \left(3 f'(r) z'(r)^2 r^3+2 d-8\right)\right)+f(r) \left(r^2 \left((4 d-15) (2 (d-2)\right.\right.\right.\right.\nonumber\\
     &\left.\left.\left.\left.(d-1) \alpha +1) r^2+2 (d-3) \alpha  \left((6 d-11) f'(r)-2 r f''(r)\right) r-2 (d-4) (d-3) (d+8) \alpha \right) z'(r)^2\right.\right.\right.\nonumber\\
     &\left.\left.\left.+2 (d-5) (d-4) (d-3) \alpha \right)\right) z(r)^5+r z'(r) \left(-4 (d-6) (d-4) (d-3) r^2 \alpha  z'(r)^2 f(r)^3+2 \left(2 (d-4) \right.\right.\right.\nonumber\\
     &\left.\left.\left.\alpha  (d-3)^2+r^2 z'(r)^2 \left((2 d-3) (2 (d-2) (d-1) \alpha +1) r^2-(d-3) \alpha  \left((3 d-14) f'(r)+2 r f''(r)\right) r\right.\right.\right.\right.\nonumber\\
     &\left.\left.\left.\left.+2 (d-4) (d-3) (4 d-37) \alpha \right)\right) f(r)^2+\left(-8 (d-4) \alpha  (d-3)^2-2 (d-1) r^2 (2 (d-2) (d-1) \alpha +1)\right.\right.\right.\nonumber\\
     &\left.\left.\left.+r \left(f'(r) \left(r^2 \left(3 (2 (d-2) (d-1) \alpha +1) r^2+2 (d-3) \alpha  f'(r) r+2 (d-3) (3 d-32) \alpha \right) z'(r)^2\right.\right.\right.\right.\right.\nonumber\\
     &\left.\left.\left.\left.\left.+4 (d-3) (3 d-10) \alpha \right)+4 (d-3) r \alpha  f''(r)\right)\right) f(r)+2 r f'(r) \left(-\left((2 (d-2) (d-1) \alpha +1) r^2\right)\right.\right.\right.\nonumber\\
     &\left.\left.\left.+2 (d-3) \alpha  f'(r) r-4 (d-4) (d-3) \alpha \right)\right) z(r)^4+2 r^2 z'(r)^2 \left(4 (d-5) (d-4) (d-3) r^2 \alpha  z'(r)^2 f(r)^3\right.\right.\nonumber\\
     &\left.\left.+\left((d-5) r^2 \left((2 (d-2) (d-1) \alpha +1) r^2+4 (d-3) \alpha  f'(r) r-2 (d-4) (d-3) \alpha \right) z'(r)^2\right.\right.\right.\nonumber\\
     &\left.\left.\left.-2 (d-4)(d-3) (d+6) \alpha \right) f(r)^2+\left(-\left((2 d-7) (2 (d-2) (d-1) \alpha +1) r^2\right)+2 (d-3) \alpha  \right.\right.\right.\nonumber\\
     &\left.\left.\left.\left(r f''(r)-3 (d-2) f'(r)\right) r+2 (d-4) (d-3) (d-1) \alpha \right) f(r)-3 (d-3) r^2 \alpha  f'(r)^2\right) z(r)^3\right. \nonumber\\
     &\left.+r^3 f(r) z'(r)^3\left(4 (d-6) (d-4) (d-3) \alpha  f(r)^2+2 \left(-\left((2 d-3) (2 (d-2) (d-1) \alpha +1) r^2\right)\right.\right.\right.\nonumber\\
     &\left.\left.\left.+(d-3) \alpha  \left((3 d-14) f'(r)+2 r f''(r)\right) r-2 (d-13) (d-4) (d-3) \alpha \right) f(r)\right.\right.\nonumber\\
     &\left.\left.-r f'(r) \left(3 (2 (d-2) (d-1) \alpha +1) r^2+2 (d-3) \alpha  f'(r) r-16 (d-3) \alpha \right)\right)z(r)^2-2 (d-4)\right. \nonumber\\
     &\left.  r^4 f(r)^2 \left((2 (d-2) (d-1) \alpha +1) r^2-2 (d-4) (d-3) \alpha +4 (d-3) \alpha \left((d-4) f(r)+r f'(r)\right)\right)   \right.\nonumber\\
     &\left.z'(r)^4 z(r)-r^5 f(r)^2\left(2 (d-2) \left(r^2 (2 (d-2) (d-1) \alpha +1)-2 (d-4) (d-3) \alpha\right)\right.\right.\nonumber\\
     &\left.\left.f(r)+r \left(r^2+2 (d-2) \left((d-1) r^2-d+3\right) \alpha \right) f'(r)\right) z'(r)^5\right)/\left(2 r^2 f(r) z(r)^2 \left(z(r)^2-1\right) \right.\nonumber\\
     &\left.\left(-6 (d-3) r \alpha  \left(z(r)^2-1\right) \left(2 (d-4) f(r)+r f'(r)\right) z'(r) z(r)+r^2 f(r) \left((2 (d-2) (d-1) \alpha +1) r^2\right.\right.\right.\nonumber\\
     &\left.\left.\left.-2 (d-4) (d-3) \alpha +4 (d-3) \alpha  \left((d-4) f(r)+r f'(r)\right)\right) z'(r)^2-z(r)^2 \left(z(r)^2-1\right) \left((2 (d-2) \right.\right.\right.\nonumber\\
     &\left.\left.\left. (d-1) \alpha+1) r^2+4 (d-4) (d-3) \alpha +2 (d-3) \alpha  \left(-\left((d-4) \left(3 z(r)^2+f(r)\right)\right)-r f'(r)\right)\right)\right)\right)
\end{align}

{\bf  EOMs for No-island phase } Here we give the EOMs for No-island phase. As there are two unknown functions $r(z)$ and $v(z)$, we have two EOMs, which are given by:
\begin{align}\label{EOMHM}
    &-2 f(r(z))^3 v'(z) \left(\left(z^2-1\right) v'(z)-2\right) \left((d-2) \left(2 \left(d^2-3 d+2\right) \alpha +1\right) v'(z)^2 \left(\left(z^2-1\right) v'(z)-2\right)^2 r(z)^4\right.\nonumber\\
    &\left.-2 \left(d^2-7 d+12\right) \alpha  \left((d-4) z^2 \left(z^2-1\right)^2 v'(z)^4-4 (d-4) z^2 \left(z^2-1\right) v'(z)^3+\left(-16 z^2+d \left(5 z^2-1\right)+2\right)\right.\right.\nonumber\\
    &\left.\left.v'(z)^2-2 \left(d+\left(z-z^3\right) v''(z)-2\right) v'(z)-2 z v''(z)\right) r(z)^2+4 \left(d^2-7 d+12\right) z \alpha  \left(z v'(z) \left(2-\left(z^2-1\right) v'(z)\right)\right.\right.\nonumber\\
    &\left.\left.r''(z)+r'(z) \left(\left(-3 z^2+d \left(z^2-1\right)+4\right) v'(z)^2+\left(-2 d+z \left(z^2-1\right) v''(z)+8\right) v'(z)-z v''(z)\right)\right) r(z)\right.\nonumber\\
    &\left.-2 \left(d^3-13 d^2+54 d-72\right) z^2 \alpha  r'(z)^2 v'(z) \left(\left(z^2-1\right) v'(z)-2\right)\right) r(z)^3+f(r(z))^2 \left(-\left(\left(2 \left(d^2-3 d+2\right) \alpha +1\right)\right.\right.\nonumber\\
    &\left.\left.f'(r(z)) v'(z)^3 \left(\left(z^2-1\right) v'(z)-2\right)^3 r(z)^7\right)+2 v'(z) \left(\left(z^2-1\right) v'(z)-2\right) \left((d-3) \alpha  f'(r(z)) \left((d-4) z^2 \left(z^2-1\right)^2\right.\right.\right.\nonumber \\
    &\left.\left.\left.v'(z)^4-4 (d-4) z^2 \left(z^2-1\right) v'(z)^3+\left(-16 z^2+d \left(5 z^2-1\right)+2\right) v'(z)^2-2 \left(d+\left(z-z^3\right) v''(z)-2\right) v'(z) \right.\right.\right.\nonumber\\
    &\left.\left.\left.-2 z v''(z)\right)-z \left(2 \left(d^2-3 d+2\right) \alpha +1\right) \left(z v'(z) \left(2-\left(z^2-1\right) v'(z)\right) r''(z)+r'(z) \left(\left(-3 z^2+d \left(z^2-1\right)+4\right) v'(z)^2\right.\right.\right.\right. \nonumber\\
    &\left.\left.\left.\left.+\left(-2 d+z \left(z^2-1\right) v''(z)+8\right) v'(z)-z v''(z)\right)\right)\right) r(z)^5+2 z v'(z) \left(\left(z^2-1\right) v'(z)-2\right) \left(-z v'(z) \left(\left(z^2-1\right) v'(z)\right.\right.\right.\nonumber\\
    &\left.\left.\left.-2\right) \left((2 d-3) \left(2 \left(d^2-3 d+2\right) \alpha +1\right)-2 (d-3) \alpha  f''(r(z))\right) r'(z)^2-4 (d-3) \alpha  f'(r(z)) \left(z v'(z) \left(2-\left(z^2-1\right) v'(z)\right)\right.\right.\right.\nonumber\\
    &\left.\left.\left.r''(z)+r'(z) \left(\left(-3 z^2+d \left(z^2-1\right)+4\right) v'(z)^2+\left(-2 d+z \left(z^2-1\right) v''(z)+8\right) v'(z)-z v''(z)\right)\right)\right)r(z)^4\right. \nonumber\\
    &\left.+2 (d-3) z \alpha  \left((3 d-14) z f'(r(z)) r'(z)^2 v'(z)^2 \left(\left(z^2-1\right) v'(z)-2\right)^2-2 (d-4) z v'(z) \left(z^2 \left(z^2-1\right)^2 v'(z)^3\right.\right.\right.\nonumber\\
    &\left.\left.\left.-4 z^2 \left(z^2-1\right) v'(z)^2+\left(5 z^2-1\right) v'(z)-2\right) r''(z)+2 (d-4) r'(z) \left(z^2 \left(z^2-1\right) \left(-5 z^2+d \left(z^2-1\right)+6\right) v'(z)^4\right.\right.\right.\nonumber\\
    &\left.\left.\left.+z^2 \left(-4 d z^2+22 z^2+\left(z^2-1\right)^2 v''(z) z+4 d-24\right) v'(z)^3+\left(-3 \left(z^2-1\right) v''(z) z^3+5 d z^2-25 z^2-d+4\right) v'(z)^2\right.\right.\right.\nonumber\\
    &\left.\left.\left.+\left(-2 d+z \left(5 z^2-3\right) v''(z)+8\right) v'(z)-3 z v''(z)\right)\right) r(z)^3+4 \left(d^2-7 d+12\right) z^2 \alpha  r'(z) \left(6 z v'(z) \left(\left(z^2-1\right) v'(z)-2\right)\right.\right.\nonumber\\
    &\left.\left.r''(z)+r'(z) \left((2 d-7) z^2 \left(z^2-1\right)^2 v'(z)^4-4 (2 d-7) z^2 \left(z^2-1\right) v'(z)^3+\left(7 d z^2-19 z^2+d-13\right) v'(z)^2\right.\right.\right.\nonumber\\
    &\left.\left.\left.+2 \left(d-2 z \left(z^2-1\right) v''(z)-13\right) v'(z)+4 z v''(z)\right)\right) r(z)^2-4 \left(d^2-7 d+12\right) z^3 \alpha  r'(z)^2 \left(z v'(z) \left(\left(z^2-1\right) v'(z)-2\right) \right.\right.\nonumber\\
    &\left.\left.r''(z)+r'(z) \left(\left(5 z^2+d \left(z^2-1\right)-6\right) v'(z)^2-\left(2 (d+6)+z \left(z^2-1\right) v''(z)\right) v'(z)+z v''(z)\right)\right) r(z)+4 (d-4) (d-3)^2\right.\nonumber\\
    &\left.z^4 \alpha  r'(z)^4 v'(z) \left(\left(z^2-1\right) v'(z)-2\right)\right) r(z)+2 z^3 r'(z)^3 \left(2 \left(d^3-12 d^2+47 d-60\right) \alpha  r'(z)^2 z^4+4 \left(d^2-7 d+12\right)\right.\nonumber\\
    &\left.\alpha  r(z) f'(r(z)) r'(z)^2 z^2+2 (d-3) \alpha  r(z)^3 f'(r(z))^2 r'(z) v'(z) \left(\left(z^2-1\right) v'(z)-2\right) z-(d-3) r(z)^2 r'(z) \left(z r'(z)\right.\right. \nonumber\\
    &\left.\left.\left(2 \left(d^2-3 d+2\right) \alpha -2 f''(r(z)) \alpha +1\right)-2 (d-4) \alpha  f'(r(z)) \left(\left(z^2-1\right) v'(z)^2 z^2-2 v'(z) z^2-2\right)\right) z-r(z)^4 f'(r(z))\right.\nonumber\\
    &\left.\left(3 (d-3) \alpha  f'(r(z))+z \left(2 \left(d^2-3 d+2\right) \alpha +1\right) r'(z)\right) v'(z) \left(\left(z^2-1\right) v'(z)-2\right)\right)+z^2 f(r(z)) r'(z) \nonumber\\
    &\left(-3 \left(2 \left(d^2-3 d+2\right) \alpha +1\right) f'(r(z)) r'(z) v'(z)^2 \left(\left(z^2-1\right) v'(z)-2\right)^2 r(z)^6-2 (d-3) \alpha  f'(r(z))^2 r'(z) v'(z)^2\right. \nonumber\\
    &\left.\left(\left(z^2-1\right) v'(z)-2\right)^2 r(z)^5+2 \left((d-3) \alpha  f'(r(z)) \left(6 z v'(z) \left(\left(z^2-1\right) v'(z)-2\right) r''(z)+r'(z) \left(3 (d-4) z^2\right.\right.\right.\right. \nonumber\\
    &\left.\left.\left.\left.\left(z^2-1\right)^2 v'(z)^4-12 (d-4) z^2 \left(z^2-1\right) v'(z)^3+4 \left((3 d-11) z^2-2\right) v'(z)^2-4 \left(z \left(z^2-1\right) v''(z)+4\right) v'(z)\right.\right.\right.\right.\nonumber\\
    &\left.\left.\left.\left.+4 z v''(z)\right)\right)+z r'(z) \left(z \left(2 \left(d^2-3 d+2\right) \alpha +1\right) v'(z) \left(\left(z^2-1\right) v'(z)-2\right) r''(z)+r'(z) \left(\left(2 (d-3) \left(z^2-1\right) \right.\right.\right.\right.\right.\nonumber\\
    &\left.\left.\left.\left.\left.\alpha  f''(r(z))-\left(-6 z^2+2 d \left(z^2-1\right)+7\right) \left(2 \left(d^2-3 d+2\right) \alpha +1\right)\right) v'(z)^2+\left(\left(2 \left(d^2-3 d+2\right) \alpha +1\right) \left(4 d+\left(z-z^3\right)\right.\right.\right.\right.\right.\right.\nonumber\\
    &\left.\left.\left.\left.\left.\left.v''(z)-14\right)-4 (d-3) \alpha  f''(r(z))\right) v'(z)+z \left(2 \left(d^2-3 d+2\right) \alpha +1\right) v''(z)\right)\right)\right) r(z)^4+2 z r'(z) \left(-z v'(z) \left(\left(z^2-1\right)\right.\right.\right.\nonumber\\
    &\left.\left.\left.v'(z)-2\right) \left((d-1) \left(2 \left(d^2-3 d+2\right) \alpha +1\right)-2 (d-3) \alpha  f''(r(z))\right) r'(z)^2-2 (d-3) \alpha  f'(r(z)) \left(z v'(z) \left(\left(z^2-1\right)\right.\right.\right.\right.\nonumber\\ 
    &\left.\left.\left.\left.v'(z)-2\right) r''(z)+r'(z) \left(\left(-7 z^2+3 d \left(z^2-1\right)+6\right) v'(z)^2+\left(\left(z-z^3\right) v''(z)-6 (d-2)\right) v'(z)+z v''(z)\right)\right)\right) r(z)^3\right.\nonumber\\
    &\left.+4 (d-3) z \alpha  r'(z) \left((3 d-10) z f'(r(z)) v'(z) \left(\left(z^2-1\right) v'(z)-2\right) r'(z)^2+(d-4) \left(-v''(z) z^3+\left(-10 z^2+2 d \left(z^2-1\right)\right.\right.\right.\right.\nonumber\\
    &\left.\left.\left.\left.+11\right) v'(z)^2 z^2+v'(z) \left(-4 d+z \left(z^2-1\right) v''(z)+22\right) z^2+d-1\right) r'(z)-(d-4) z \left(\left(z^2-1\right) v'(z)^2 z^2-2 v'(z) z^2-2\right)\right.\right.\nonumber\\
    &\left.\left.r''(z)\right) r(z)^2+4 (d-4) (d-3)^2 z^2 \alpha  r'(z)^3 \left(\left(z^2-1\right) v'(z)^2 z^2-2 v'(z) z^2-2\right) r(z)\right.\nonumber\\
    &\left.+4 \left(d^3-12 d^2+47 d-60\right) z^3 \alpha  r'(z)^4\right)=0\nonumber
\end{align}
and 
\begin{align}
    &-2 z^4 r(z) f'(r(z)) \left(-2 \left(d^2-7 d+12\right) \alpha  r'(z) z^3-2 (d-3) \alpha  r(z) f'(r(z)) r'(z) z+r(z)^2 \left(2 (d-3) \alpha  f'(r(z))\right.\right.\nonumber\\
    &\left.\left.+z \left(2 \left(d^2-3 d+2\right) \alpha +1\right) r'(z)\right)\right) \left(\left(z^2-1\right) v'(z)-1\right) r'(z)^4+z^2 f(r(z)) \left(4 (d-4) (d-3)^2 \alpha  r'(z)^3 \left(\left(z^2-1\right) v'(z)\right.\right.\nonumber\\
    &\left.\left.-1\right) z^5+4 (d-3) \alpha  r(z) r'(z) \left(-\left((d-4) \left(\left(z^2-1\right) v'(z)-1\right) r''(z) z^2\right)-(d-4) r'(z) \left(-d+\left(-5 z^2+d \left(z^2-1\right)+3\right)\right.\right.\right.\nonumber\\
    &\left.\left.\left.v'(z)+\left(z-z^3\right) v''(z)+3\right) z+(3 d-10) f'(r(z)) r'(z)^2 \left(\left(z^2-1\right) \right.\right.\right.\nonumber\\
    &\left.\left.\left.v'(z)-1\right)\right) z^3+2 r(z)^2 r'(z) \left(-z \left(\left(z^2-1\right) v'(z)-1\right) \left((d-1) \left(2 \left(d^2-3 d+2\right) \alpha +1\right)-2 (d-3) \alpha  f''(r(z))\right) r'(z)^2\right.\right.\nonumber\\
    &\left.\left.-2 (d-3) \alpha  f'(r(z)) \left(z \left(\left(z^2-1\right) v'(z)-1\right) r''(z)+r'(z) \left(-6 (d-3)+2 \left(-10 z^2+3 d \left(z^2-1\right)+9\right) v'(z)\right.\right.\right.\right.\nonumber\\
    &\left.\left.\left.\left.+\pqty{z-z^3} v''(z)\right)\right)\right) z^2-2 (d-3) \alpha  r(z)^4 f'(r(z))^2 r'(z) v'(z) \left((\pqty{z^2-1}^2 v'(z)^2-3 \pqty{z^2-1} v'(z)+2\right) z\right.\notag\\
    &\left.+2 r(z)^3 \left((d-3) \alpha  f'(r(z)) \left(4 z \left(\pqty{z^2-1} v'(z)-1\right) r''(z)+r'(z) \left(3 (d-4) z^2 \pqty{z^2-1}^2 v'(z)^3\right.\right.\right.\right.\notag\\
    &\left.\left.\left.\left.-9 (d-4) z^2 \pqty{z^2-1} v'(z)^2+2 \left(-22 z^2+d \pqty{6 z^2-3}+8\right) v'(z)-6 d-2 z \pqty{z^2-1} v''(z)+16\right)\right)\right.\right.\notag\\
    &\left.\left.+z r'(z) \left(z \pqty{2 \pqty{d^2-3 d+2} \alpha +1} \pqty{\pqty{z^2-1} v'(z)-1} r''(z)+r'(z) \left(2 (d-3) \alpha  f''(r(z))\right.\right.\right.\right.\notag\\
    &\left.\left.\left.\left.+v'(z) \left(\pqty{-3 z^2+d \pqty{z^2-1}+1}\left(2 \pqty{d^2-3 d+2} \alpha +1\right)-2 (d-3) \pqty{z^2-1} \alpha  f''(r(z))\right)\right.\right.\right.\right.\notag\\
    &\left.\left.\left.\left.-\pqty{2 \pqty{d^2-3 d+2} \alpha +1} \pqty{d+z \pqty{z^2-1} v''(z)-1}\right)\right)\right) z+r(z)^5 f'(r(z)) \left(2 (d-3) \alpha  f'(r(z))\right.\right.\notag\\
    &\left.\left.-3 z \pqty{2 \pqty{d^2-3 d+2} \alpha +1} r'(z)\right) v'(z) \left(\pqty{z^2-1}^2 v'(z)^2-3 \pqty{z^2-1} v'(z)+2\right)\right) r'(z)^2\notag\\
    &+2 f(r(z))^3 r(z)^2 \left(\pqty{2 \pqty{d^2-3 d+2} \alpha +1} v'(z) \left(\pqty{z^2-1} v'(z)-2\right) \left(\pqty{z^2-1} \pqty{-3 z^2+d \pqty{z^2-1}+2} v'(z)^3\right.\right.\notag\\
    &\left.\left.+\pqty{9 z^2-3 d \pqty{z^2-1}-6} v'(z)^2+2 (d-2) v'(z)+z v''(z)\right) r(z)^5-(d-2) z \pqty{2 \pqty{d^2-3 d+2} \alpha +1} r'(z) v'(z)^2 \left(\left(z^2\right.\right.\right.\notag\\
    &\left.\left.\left.-1\right) v'(z)-2\right)^2 \pqty{\pqty{z^2-1} v'(z)-1} r(z)^4-2 \pqty{d^2-7 d+12} \alpha  \left(z^2 \pqty{z^2-1}^2 \left(-5 z^2+d \left(z^2-1\right)+4\right) v'(z)^5\right.\right.\notag\\
    &\left.\left.-5 z^2 \left(z^2-1\right) \left(-5 z^2+d \left(z^2-1\right)+4\right) v'(z)^4+\left(9 d z^4-39 z^4-10 d z^2+35 z^2+d-2\right) v'(z)^3\right.\right.\notag\\
    &\left.\left.+\left(23 z^2+\left(3 z^4-5 z^2+2\right) v''(z) z+d \left(3-7 z^2\right)-6\right) v'(z)^2+2 \left(d+\left(2 z-3 z^3\right) v''(z)-2\right) v'(z)+3 z v''(z)\right) r(z)^3\right.\notag\\
    &\left.+2 \left(d^2-7 d+12\right) z \alpha  \left(r'(z) \left((d-4) z^2 \left(z^2-1\right)^3 v'(z)^5-5 (d-4) z^2 \left(z^2-1\right)^2 v'(z)^4+\left(z^2-1\right) \left(-40 z^2+d \left(11 z^2\right.\right.\right.\right.\right.\notag\\
    &\left.\left.\left.\left.\left.-3\right)+10\right) v'(z)^3+\left(-13 d z^2+48 z^2+4 \left(z^2-1\right)^2 v''(z) z+9 d-30\right) v'(z)^2+\left(6 d-8 z \left(z^2-1\right) v''(z)-20\right) v'(z)\right.\right.\right.\notag\\
    &\left.\left.\left.+6 z v''(z)\right)-2 z v'(z) \left(\left(z^2-1\right)^2 v'(z)^2-3 \left(z^2-1\right) v'(z)+2\right) r''(z)\right) r(z)^2-2 \left(d^2-7 d+12\right) z^2 \alpha  r'(z) \left(r'(z) \left(\left(z^2 \right.\right.\right.\right.\notag\\
    &\left.\left.\left.\left.-1\right) \left(-13 z^2+3 d \left(z^2-1\right)+14\right) v'(z)^3+\left(-9 d z^2+43 z^2+2 \left(z^2-1\right)^2 v''(z) z+9 d-42\right) v'(z)^2\right.\right.\right.\notag\\
    &\left.\left.\left.+\left(6 d-4 z \left(z^2-1\right) v''(z)-28\right) v'(z)+3 z v''(z)\right)-2 z v'(z) \left(\left(z^2-1\right)^2 v'(z)^2-3 \left(z^2-1\right) v'(z)+2\right) r''(z)\right) r(z)\right.\notag\\
    &\left.+2 \left(d^3-13 d^2+54 d-72\right) z^3 \alpha  r'(z)^3 v'(z) \left(\left(z^2-1\right)^2 v'(z)^2-3 \left(z^2-1\right) v'(z)+2\right)\right)\notag\\
    &+z f(r(z))^2 \left(-\left(\left(2 \left(d^2-3 d+2\right) \alpha +1\right) f'(r(z)) r'(z) v'(z)^2 \left(\left(z^2-1\right) v'(z)-2\right)^2 \left(\left(z^2-1\right) v'(z)-1\right) r(z)^7\right)\right.\notag\\
    &\left.+2 \left((d-3) \alpha  f'(r(z)) \left(r'(z) \left((d-4) z^2 \left(z^2-1\right)^3 v'(z)^5-5 (d-4) z^2 \left(z^2-1\right)^2 v'(z)^4+\left(z^2-1\right) \left(-40 z^2\right.\right.\right.\right.\right.\notag\\
    &\left.\left.\left.\left.\left.+d \left(11 z^2-3\right)+10\right) v'(z)^3+\left(-13 d z^2+48 z^2+4 \left(z^2-1\right)^2 v''(z) z+9 d-30\right) v'(z)^2+\left(6 d-8 z \left(z^2-1\right) v''(z)\right.\right.\right.\right.\right.\notag\\
    &\left.\left.\left.\left.\left.-20\right) v'(z)+6 z v''(z)\right)-2 z v'(z) \left(\left(z^2-1\right)^2 v'(z)^2-3 \left(z^2-1\right) v'(z)+2\right) r''(z)\right)+z r'(z) \left(z \left(2 \left(d^2-3 d+2\right) \alpha \right.\right.\right.\right.\notag\\
    &\left.\left.\left.\left.+1\right) v'(z) \left(\left(z^2-1\right)^2 v'(z)^2-3 \left(z^2-1\right) v'(z)+2\right) r''(z)+r'(z) \left(\left(z^2-1\right) \left(\left(-6 z^2+2 d \left(z^2-1\right)+3\right) \left(2 \left(d^2-3 d\right.\right.\right.\right.\right.\right.\right.\notag\\
    &\left.\left.\left.\left.\left.\left.\left.+2\right) \alpha +1\right)-2 (d-3) \left(z^2-1\right) \alpha  f''(r(z))\right) v'(z)^3-\left(\left(2 \left(d^2-3 d+2\right) \alpha +1\right) \left(-16 z^2+\left(z^2-1\right)^2 v''(z) z+6 d \left(z^2 \right.\right.\right.\right.\right.\right.\right.\notag\\
    &\left.\left.\left.\left.\left.\left.\left.-1\right)+9\right)-6 (d-3) \left(z^2-1\right) \alpha  f''(r(z))\right) v'(z)^2+2 \left(\left(2 \left(d^2-3 d+2\right) \alpha +1\right) \left(2 d+z \left(z^2-1\right) v''(z)-3\right)\right.\right.\right.\right.\right.\notag\\
    &\left.\left.\left.\left.\left.-2 (d-3) \alpha  f''(r(z))\right) v'(z)+z \left(2 \left(d^2-3 d+2\right) \alpha +1\right) v''(z)\right)\right)\right) r(z)^5-2 z r'(z) \left(z v'(z) \left(\left(z^2-1\right)^2 v'(z)^2 \right.\right.\right.\notag\\
    &\left.\left.\left.-3 \left(z^2-1\right) v'(z)+2\right) \left((2 d-3) \left(2 \left(d^2-3 d+2\right) \alpha +1\right)-2 (d-3) \alpha  f''(r(z))\right) r'(z)^2+2 (d \right.\right.\notag\\
    &\left.\left.-3) \alpha  f'(r(z)) \left(r'(z) \left(\left(z^2-1\right) \left(-11 z^2+3 d \left(z^2-1\right)+12\right) v'(z)^3+\left(-9 d z^2+37 z^2+2 \left(z^2-1\right)^2 v''(z) z+9 d\right.\right.\right.\right.\right.\notag\\
    &\left.\left.\left.\left.\left.-36\right) v'(z)^2+\left(6 (d-4)-4 z \left(z^2-1\right) v''(z)\right) v'(z)+3 z v''(z)\right)-2 z v'(z) \left(\left(z^2-1\right)^2 v'(z)^2-3 \left(z^2-1\right) v'(z) \right.\right.\right.\right.\notag\\
    &\left.\left.\left.\left.+2\right) r''(z)\right)\right) r(z)^4-2 (d-3) z \alpha  r'(z) \left(-\left((3 d-14) z f'(r(z)) v'(z) \left(\left(z^2-1\right)^2 v'(z)^2-3 \left(z^2-1\right) v'(z)+2\right) r'(z)^2\right) \right.\right.\notag\\
    &\left.\left.+2 (d-4) \left(z^2 \left(z^2-1\right) \left(-10 z^2+2 d \left(z^2-1\right)+7\right) v'(z)^3-z^2 \left(-28 z^2+\left(z^2-1\right)^2 v''(z) z+6 d \left(z^2-1\right)+21\right) v'(z)^2 \right.\right.\right.\notag\\
    &\left.\left.\left.+\left(2 \left(z^2-1\right) v''(z) z^3-15 z^2+d \left(5 z^2-1\right)-1\right) v'(z)-d+z v''(z)-1\right) r'(z)+2 (d-4) z \left(z^2 \left(z^2-1\right)^2 v'(z)^3 \right.\right.\right.\notag\\
    &\left.\left.\left.-3 z^2 \left(z^2-1\right) v'(z)^2+\left(5 z^2-3\right) v'(z)-3\right) r''(z)\right) r(z)^3+4 \left(d^2-7 d+12\right) z^2 \alpha  r'(z)^2 \left(4 z \left(\left(z^2-1\right) v'(z)-1\right) r''(z) \right.\right.\notag\\
    &\left.\left.+r'(z) \left((2 d-7) z^2 \left(z^2-1\right)^2 v'(z)^3-3 (2 d-7) z^2 \left(z^2-1\right) v'(z)^2+\left(-19 z^2+d \left(7 z^2-3\right)+1\right) v'(z)-3 d-2 z \left(z^2 \right.\right.\right.\right.\notag\\
    &\left.\left.\left.\left.-1\right) v''(z)+1\right)\right) r(z)^2-4 \left(d^2-7 d+12\right) z^3 \alpha  r'(z)^3 \left(z \left(\left(z^2-1\right) v'(z)-1\right) r''(z)+r'(z) \left(-3 d+\left(-7 z^2+3 d \left(z^2-1\right) \right.\right.\right.\right.\notag\\
    &\left.\left.\left.\left.+5\right) v'(z)+\left(z-z^3\right) v''(z)+5\right)\right) r(z)+4 (d-4) (d-3)^2 z^4 \alpha  r'(z)^5 \left(\left(z^2-1\right) v'(z)-1\right)\right)=0
\end{align}

\end{document}